\documentclass[numberedappendix]{emulateapj}
\usepackage{amsmath}

\slugcomment{accepted by ApJ; \today}

\newcommand{\G}{\Gamma}

\newcommand{\sT}{\sigma_{\rm T}}
\newcommand{\p}{^\prime}
\newcommand{\e}{\epsilon}
\newcommand{\g}{\gamma}
\newcommand{\gp}{\gamma^{\prime}}

\newcommand{\tp}{t^\prime}

\newcommand{\dD}{\delta_{\rm D}}

\newcommand{\psim}{\lower.5ex\hbox{$\; \buildrel \propto \over\sim \;$}}
\newcommand{\lbar}{\lower.0ex\hbox{$\; \buildrel
{\lower0.0ex \hbox{-}} \over\lambda  \;$}}

\newcommand{\dotg}{\dot{\g}}
\newcommand{\tilN}{\tilde{N}}
\newcommand{\tilQ}{\tilde{Q}}
\newcommand{\tilF}{\tilde{F}}

\newcommand{\cm}{\mathrm{cm}}

\newcommand{\erg}{\mathrm{erg}}

\newcommand{\keV}{\mathrm{keV}}

\newcommand{\GeV}{\mathrm{GeV}}

\newcommand{\s}{\mathrm{s}}

\newcommand{\Gauss}{\mathrm{G}}

\shorttitle{Fourier Analysis of Blazar Variability}
\shortauthors{Finke \& Becker}

\begin{document}
\title{Fourier Analysis of Blazar Variability:  Klein-Nishina Effects and the 
Jet Scattering Environment}

\author{Justin D.\ Finke}
\affil{U.S.\ Naval Research Laboratory, Code 7653, 4555 Overlook Ave.\ SW,
        Washington, DC,
        20375-5352}
\email{justin.finke@nrl.navy.mil}
\author{Peter A.\ Becker}
\affil{	School of Physics, Astronomy, and Computational Sciences, MS 5C3, 
	George Mason University, 4400 University Drive, Fairfax, VA 22030}
\email{pbecker@gmu.edu}



\begin{abstract}

The strong variability of blazars can be characterized by power
spectral densities (PSDs) and Fourier frequency-dependent time lags.
In previous work, we created a new theoretical formalism for
describing the PSDs and time lags produced via a combination of
stochastic particle injection and emission via the synchrotron,
synchrotron self-Compton, and external Compton (EC) processes.  This
formalism used the Thomson cross section and simple $\delta$-function
approximations to model the synchrotron and Compton emissivities.
Here we expand upon this work, using the full Compton cross section
and detailed and accurate emissivities.  Our results indicate good
agreement between the PSDs computed using the $\delta$-function
approximations and those computed using the accurate expressions,
provided the observed photons are produced primarily by electrons with
energies exceeding the lower limit of the injected particle
population. Breaks are found in the PSDs at frequencies corresponding
to the cooling timescales of the electrons primarily responsible for
the observed emission, and the associated time lags are related to the
difference in electron cooling timescales between the two energy
channels, as expected. If the electron cooling timescales can be
determined from the observed time lags and/or the observed EC PSDs,
then one could in principle use the method developed here to determine
the energy of the external seed photon source for EC, which is an
important unsolved problem in blazar physics.

\end{abstract}


\section{Introduction}
\label{intro}

Blazars, active galactic nuclei (AGN) with jets aligned with our line
of sight, relativistically beam nonthermal radiation towards the Earth.
This emission extends across the electromagnetic spectrum, from radio
to $\gamma$ rays.  Although the observed radio emission probably
comes from the superposition of many self-absorbed synchrotron
components spread out over several parsecs \citep{konigl81}, the
location along the jet where the higher-frequency emission originates
is a matter of some controversy.  This is particularly true for the
most powerful class of blazars, the flat spectrum radio quasars
(FSRQs) with strong broad emission lines.  These objects are thought
to emit $\gamma$-rays primarily through Compton scattering of an
external radiation field (known as external Compton or EC).  However,
the primary external radiation field for EC is not known, and could be
provided by the accretion disk \citep{dermer93,dermer02}, the broad line region
\citep[BLR;][]{sikora94}, or the dust torus
\citep[][]{kataoka99,blazejowski00}.  This uncertainty in the seed
photon source reflects the uncertainty in the location along the jet
of the primary $\gamma$-ray emitting region, which one would expect to
be within $\sim 0.1$\ pc from the black hole (BH) if the scattering of
disk radiation dominates, or between $\sim 0.1$ and $\sim 1$\ pc if
the scattering of BLR emission dominates, or $\ga 1$\ pc if the
scattering of dust torus emission dominates
\citep{ghisellini09_canonical,sikora09}.

The association of $\gamma$-ray flares with the ejection of
superluminal components \citep[e.g.,][]{marscher12} indicates that the
flares are coincident with the 43 GHz core, which is likely located at
$\ga$ a few pc from the black hole, outside the BLR \citep[although
see][]{nalewajko14}.  The detection of $\ga 100\ \GeV$ $\gamma$ rays
from FSRQs also indicates the $\gamma$ rays must originate from inside
the BLR to avoid $\g\g$ absorption with BLR photons
\citep[e.g.,][]{aleksic11}.  However, rapid $\gamma$-ray variability
observed in FSRQs such as 3C 454.3 \citep{tavecchio10}, PKS 1510$-$089
\citep{brown13,saito13} and 4C 21.35 \citep[also known as PKS
1222+21;][]{aleksic11} limits the size of the emitting region, and,
assuming the emitting region takes up the entire cross section of a
conical jet, implies that it should be within $\sim 0.1$ pc of the
black hole, and hence inside the BLR.  \citet{dotson12} point out a
way to distinguish between these scenarios based on the different
cooling timescales at two energies if the seed photons are from the
BLR (e.g. with dimensionless energy for Ly$\alpha$ $\e_0 = E_0/m_ec^2
= 2\times10^{-5}$) or from the dust torus (e.g. $\sim1000$\ K dust
producing photons with dimensionless energy $\e_0 = 5\times10^{-7}$)
due to the variation of the Compton cross-section at high energies.

Blazars are bright and highly variable at all wavelengths.  Their
variability appears stochastic, characterized by power-law colored
noise in their power spectral densities (PSDs) at all wavelengths,
including radio \citep[e.g.,][]{trippe11,park14}, optical
\citep[e.g.,][]{chatterjee12,wehrle13,edelson13,revalski14}, X-rays
\citep[e.g.,][]{zhang99,kataoka01,zhang02_mrk421}, and $\gamma$ rays
\citep[e.g.,][]{aharonian07_2155,abdo10_var,nakagawa13}.  Despite the
popularity of PSDs for characterizing blazar variability, until
recently a theoretical motivation has been lacking.  Previously, we
presented a new theoretical formalism for the modeling and
interpretation of these sources \citep[][hereafter Paper~I]{finke14}.
This work was based on solving the Fourier-transformed electron
transport (continuity) equation for the case of synchrotron and
Thomson cooling, yielding a solution for the electron distribution as
a function of Fourier frequency.  This solution was then combined with
simple $\delta$-function approximations for synchrotron and Compton
scattering emission to predict the expected PSDs and Fourier
frequency-dependent time lags one would expect to observe from
blazars.  We showed that these Fourier transform-related data products
contain detailed information about the characteristic timescales for
particle escape, energy losses, and light crossing in the emission
region, which is taken to be an outflowing spherical plasma blob.  
Several of these features were also noted by \citet{mastich13}.

Here we extend the work presented in Paper~I, by solving the electron
continuity equation with the full Compton cooling rates, including
Klein-Nishina effects (Section \ref{electrondist}).  We then compute
the synchrotron and EC PSDs, going beyond the simple $\delta$-function
approximations adopted in Paper I and employing well-known precise
expressions for the synchrotron emissivity and Compton scattering
cross section (Section \ref{observedpsd}).  We also improve on our
previous work by incorporating the light travel time effects specific
to a spherical geometry, as described by \citet{zacharias13}.  We do
this in the context of EC emission from FSRQs.  In Section
\ref{timelag} we compute the associated time lags, and in Section
\ref{seedobtain} we describe how the electron cooling timescales,
obtained from spectral breaks of PSDs, can be used to constrain the
seed photon energy of the external radiation source for EC.  This
method is very close to the one outlined by \citet{dotson12}, except
that we suggest using PSDs to obtain the cooling timescales, rather
than directly observing the decaying part of the flare light curve.

Note that this paper deals with the PSDs and time lags from nonthermal
emission from relativistic jets.  The PSDs from coronae in Galactic
black holes and AGN exhibit breaks correlated with the mass of the
black hole \citep{mchardy06}.  This is true for the emission from the
coronae of AGN with jets as well
\citep[e.g.,][]{chatterjee09,chatterjee11}.  There is some
  theoretical indication that coronal emission, reflected off of an
  accretion disk, can create Fourier frequency-dependent time lags
  that may be related to black hole mass \citep{emman14}.  However, in
  this paper, we consider only nonthermal jet emission, beamed away
  from the jet, so that no such features are expected.
  There is no expected relation between black hole mass and PSDs and
  time lags in our model.  Although there may be a connection
between jet and coronal emission, we do not consider such an effect
here.

\section{Electron Distribution in the Fourier Frequency Domain}
\label{electrondist}

Consider a spherical homogeneous ``blob'' of Thomson-thin plasma of
radius $R$ containing nonthermal electrons and positrons (both of
which are hereafter referred to as electrons) at highly relativistic
energies, and a tangled magnetic field of strength $B$.  The electron
distribution is assumed to be isotropic and the number of electrons
between Lorentz factors $\g$ and $\g+d\g$ as a function of time $t$ is
$N_e(\g;t)$.  Electrons may be injected into the blob at a rate
$Q(\g,t)$, lose or gain energy at a rate $\dotg(\g,t)$, and escape
from the blob on a timescale $t_{\rm esc}(\g,t)$.  In this case, the
evolution of $N_e(\g;t)$ is described by
\citep[e.g.,][]{mast95,chiaberge99,li00,boett02a,chen11,chen12}
\begin{flalign}
\label{conteqn}
\frac{\partial N_e}{\partial t} + \frac{\partial}{\partial\g}[ \dotg(\g,t) N_e(\g; t) ] +
\frac{N_e(\g; t)}{t_{\rm esc}(\g,t)} 
= Q(\g, t) \ .
\end{flalign}
We will further assume that $\dotg$, $t_{\rm esc}$, and $R$ are
independent of $t$.  In this case, taking the Fourier transform of
both sides of Equation (\ref{conteqn}) gives
\begin{eqnarray}
\label{fourier_cont_eqn}
-2\pi if \tilN_e(\g,f) + \frac{\partial}{\partial\g} [\dotg(\g) \tilN_e(\g,f)] + 
\frac{\tilN_e(\g,f)}{t_{\rm esc}(\g)} 
\nonumber \\
 = \tilQ(\g, f)\ ,
\end{eqnarray}
where the tilde refers to the Fourier transform\footnote{See Paper~I
for the definitions of the Fourier transform and its inverse used
here.}, $f$ is the Fourier frequency, $\omega = 2\pi f$ is the angular
Fourier frequency, and $\tilQ(\g, f)$ is the Fourier transformed
source term.  It was shown in Paper~I that if $\dotg\le 0$ and $t_{\rm
esc}$ is independent of $\g$ then the solution to Equation
(\ref{fourier_cont_eqn}) is
\begin{flalign}
\label{Nsoln1}
\tilN_e(\g,f) & = \frac{1}{|\dotg(\g)|}\int_\g^{\infty} d\gp\ \tilQ(\gp,f)\ 
\nonumber \\ & \times
\exp\left[ -\left(\frac{1}{t_{\rm esc}} - i\omega\right)
\int_{\g}^{\gp} \frac{d\g^{\prime\prime}}{|\dotg(\g^{\prime\prime})|}
\right]\ .
\end{flalign}
We make the {\em ansatz} that the Fourier-transformed source terms is
\begin{eqnarray}
\label{tilQ}
\tilQ(\g,f) = Q_0 (f/f_0)^{-a/2}\g^{-q}H(\g;\g_1,\g_2) H(f; f_1, f_2)\ ,
\end{eqnarray}
where 
\begin{equation}
H(x; a, b) = \left\{ \begin{array}{ll}
1 & a < x <b \\
 0 & \mathrm{otherwise}
\end{array}
\right. \ 
\end{equation}
is the boxcar function.  We showed in Paper~I that the normalization constant 
\begin{flalign}
Q_0 = \frac{ 2\pi \Delta t \langle L_{\rm inj}\rangle}{m_ec^2 G\sqrt{I_r^2 + I_i^2 - 2 I_rI_0 + I_0^2}}
\end{flalign}
where $G$, $I_r$, $I_i$, and $I_0$ are integrals defined in Appendix B
of Paper~I, and $\langle L_{\rm inj}\rangle$ is the root mean squared
power injected in electrons over a time interval $\Delta t$, so that
$\Delta t$ is the duration of the light curve.

We will take into account two energy loss (i.e., cooling) processes
for $\dotg(\g)$: synchrotron and EC.  We neglect adiabatic losses,
since we assume $R$ is independent of $t$.  The synchrotron cooling
rate is given by
\begin{flalign}
-\dotg_{sy}(\g) = \frac{4c\sT}{3m_e c^2}  u_B \g^2
\label{gammadot}
\end{flalign}
where
\begin{flalign}
u_B = \frac{B^2}{8\pi}\ ,
\end{flalign}
is the magnetic energy density, $c$ is the speed of light, $m_e$ is
the electron mass, and $\sT$ is the Thomson cross section.  The
Compton cooling rate, valid in the Thomson through Klein-Nishina
regimes, for the case of scattering of an external, isotropic,
monochromatic radiation field, is given by
\begin{flalign}
-\dotg_{EC}(\g) = \frac{3c\sT u_{0}}{8m_e c^2 \e_0^2} G_{BMS}(\g\e_0)
\end{flalign}
where $u_0$ and $\e_0$ are the energy density and dimensionless photon energy, respectively, 
of the external radiation field measured in the frame of the host galaxy, and
\begin{flalign}
G_{BMS}(x) & = \frac{8}{3}x\frac{1+5x}{(1+4x)^2}\ -\ 
	\frac{4x}{1+4x}\left(\frac{2}{3}+\frac{1}{2x}+\frac{1}{8x^2}\right)
\nonumber \\
	& +\ \ln(1+4x)
\nonumber \\ & \times
\left( 1+\frac{3}{x}+\frac{3}{4}\frac{1}{x^2} + 
	\frac{\ln[1+4x]}{2x} - \frac{\ln[4x]}{x} \right)
\nonumber \\
& -\ \frac{5}{2}\frac{1}{x}\ +\ \frac{1}{x}\sum^{\infty}_{n=1}
  \frac{(1+4x)^{-n}}{n^2}\ -\ \frac{\pi^2}{6x}\ -\ 2
\end{flalign}
\citep*{boett97}.  A simple and fairly accurate approximation 
for the Compton cooling rate is
given by \citet{moderski05}, and it is
\begin{flalign}
-\dotg_{EC,M}(\g) = \frac{4c\sT}{3m_e c^2} u_{0} \g^2 M_0(4\g\e_0)
\end{flalign}
where
\begin{eqnarray}
M_0(x) = \frac{1}{(1+x)^{3/2}}\ .
\end{eqnarray}

Naturally, the total cooling rate from synchrotron and EC losses is
\begin{flalign}
\dotg(\g) = \dotg_{sy}(\g) + \dotg_{EC}(\g)\ .
\end{flalign}
We neglect the effects of SSC cooling, which can be quite complicated 
\citep{schlickeiser09,schlickeiser10,zacharias10,zacharias12,zacharias12_EC,zacharias13,zacharias14}.

\section{Observed PSDs}
\label{observedpsd}

Now consider the case where the plasma blob described in Section
\ref{electrondist} is moving with small angle $\theta\ll 1$ to our
line of sight at a speed that is a fraction $\beta$ that of light (in
the host galaxy frame) giving it a Lorentz factor
$\G=(1-\beta^2)^{-1/2}$ and a Doppler factor $\dD=[\G(1-\beta
\cos\theta)]^{-1}$.  The plasma blob comes from a supermassive black
hole in a galaxy at a cosmological distance with cosmological redshift
$z$.  Primed quantities refer to the frame co-moving with the blob,
while unprimed quantities refer to the observer's frame, except for
the quantities related to the external radiation field, $\e_0$ and
$u_0$ which are in the frame of the host galaxy, so that $\e\p_0 =
\G\e_0$ and $u\p_{0} \approx \G^2u_0$.

Below we describe the observed $\nu F_\nu$ flux in the time domain
($F(\e,t)$) and in the Fourier frequency domain ($\tilF(\e,f)$) for
synchrotron and EC for the simple $\delta$-function approximations
and for the more precise expressions, and compare them with each
other.  This uses the result for $\tilN_e(\g,f)$ described in Section
\ref{electrondist}.  The PSD can be found from the Fourier transformed
flux by
\begin{flalign}
S(\e,f) = |\tilF(\e,f)|^2 = \tilF(\e,f) \tilF^*(\e,f)\ 
\end{flalign}
where the asterisk denotes the complex conjugate.  In Paper~I, we
calculated the synchrotron emission including light travel time
effects in a very simple, cylindrical geometry.  Here, we include
light travel time effects consistent with spherical geometry, using
the results of \citet{zacharias13}.  See Appendix \ref{delayappendix}
for details.

\subsection{Synchrotron}
\label{synchsection}

\subsubsection{Delta Function Approximation}
\label{syndeltaapprox}

With the $\delta$-function approximation, synchrotron photons with
observed energy $\epsilon$ are created by electrons with a comoving
Lorentz factor given by
\begin{equation}
\label{gamma_sy}
\gp_{sy} = \sqrt{\frac{\e(1+z)}{\delta_D\e_B}}\ ,
\end{equation}
with corresponding total radiated power per electron
(cf. Equation~(\ref{gammadot}))
\begin{equation}
\dot P_e = -\dotg_{sy}(\g) m_e c^2= \frac{4c\sT}{3}  u_B \g^2 \ .
\label{Pe}
\end{equation}
The corresponding $\nu F_{\nu}$\ synchrotron flux as a function of
time $t$ (including light travel time effects) if the blob has a
tangled comoving magnetic field $B$ is
\begin{flalign}
F^{sy,\delta}(\e,t) & = \frac{6K_{sy,\delta}(1+z)}{\dD t_{lc}} \int^{2R\p/c}_{0} d\tp 
\left[ \frac{\tp c}{2R\p} - \left(\frac{\tp c}{2R\p}\right)^2 \right]
\nonumber \\ & \times
N_e\left(\gp_{sy}; \frac{t\delta_D}{1+z}-\tp\right)
\end{flalign}
where $N_e$ is the comoving electron distribution,
\begin{eqnarray}
t_{lc} = \frac{2 R\p (1+z) }{c \dD}\ 
\end{eqnarray}
denotes the light-crossing time for the spherical blob, 
\begin{flalign}
 K_{sy,\delta} = \frac{\delta_D^4}{6\pi d_L^2} c \sT u_B \gamma_{sy}^{\prime 3}\ ,
\end{flalign}
$\e_B = B/B_{cr}$, and $B_{cr} = 4.414\times10^{13}$\ G.

Similar to Paper~I, the Fourier transform of this is
\begin{flalign}
\tilF^{sy,\delta}(\e,f) & = \frac{6K_{sy,\delta}(1+z)}{2\pi if t_{lc}\dD} 
\tilN_e\left(\gp_{sy}, \frac{(1+z)f}{\dD}\right)\ 
\nonumber \\ & \times 
\left\{ \exp\left[ \frac{4 \pi i f(1+z) R^\prime}{c\dD}\right] - 1 \right\}\ .
\end{flalign}

\subsubsection{Exact Synchrotron Emissivity}
\label{synexact}

The $\nu F_{\nu}$ synchrotron  flux as a function of time $t$ including 
light travel time effects is given by
\begin{flalign}
F^{sy}(\e,t) & = \frac{6K_{sy}(1+z)}{\dD t_{lc}} \int^{2R\p/c}_{0} d\tp 
\left[ \frac{\tp c}{2R\p} - \left(\frac{\tp c}{2R\p}\right)^2 \right]
\nonumber \\ & \times 
\int_1^\infty d\gp N_e\p\left( \gp; \frac{t\dD}{1+z}-\tp \right) 
R_{CS}\left(\frac{2\e(1+z)}{3\e_B\dD\g^{\p 2}} \right)
\end{flalign}
where
\begin{eqnarray}
K_{sy}  = \frac{\sqrt{3}\dD^3 \e(1+z)e^3 B}{4\pi h d_L^2}
\end{eqnarray}
and 
\begin{flalign}
R_{CS}(x) = \frac{x}{2}\int^\pi_0 d\theta\ \sin\theta\ \int^\infty_{x/\sin\theta}dy\ K_{5/3}(y)
\end{flalign}
is the function from \citet{crusius86}. The corresponding comoving
power is again given by Equation~(\ref{Pe}).  Accurate approximations
to $R_{CS}(x)$ are given by \citet{zirak07}, \citet*{finke08_SSC}, and
\citet{joshi11}.  Following the procedure outlined in Paper~I, the
Fourier transformed synchrotron flux using this precise expression is
given by
\begin{flalign}
\tilF^{sy}(\e,f) & = \frac{6K_{sy}(1+z)}{2\pi if t_{lc}\dD}
\left\{ \exp\left[ \frac{4 \pi i f(1+z) R^\prime}{c\dD}\right] - 1 \right\}\ 
\nonumber \\ & \times 
\int_1^{\infty} d\gp \tilN_e\p \left(\gp; \frac{(1+z)f}{\dD}\right) 
R_{CS}\left(\frac{2\e(1+z)}{3\e_B\dD\g^{\p 2}} \right)\ .
\end{flalign}

\subsection{External Compton}
\label{comptonsection}

\subsubsection{Improved Delta Function Approximation}
\label{ECdeltaapprox}

In Paper~I we used a $\delta$-function approximation for Compton
scattering that was only valid in the Thomson regime.  Here we make
use of a very useful $\delta$-approximation for Compton scattering by
\citet{moderski05} that is fairly accurate (to within about 10\%) in
the Thomson ($4\dD\gp_{EC}\e_0\ll 1$) through extreme Klein-Nishina
(KN; $4\dD\gp_{EC}\e_0\gg 1$) regimes.  We assume the external radiation
field is monochromatic with dimensionless energy $\e_0$ that is
isotropic in the frame of the host galaxy.  Including light travel
time effects, this yields for the EC flux as a function of time
\begin{flalign}
\label{flux_EC_delta}
F^{EC,\delta}(\e,t) & = \frac{6K_{EC,\delta}(1+z)}{\dD t_{lc}} 
\nonumber \\ & \times 
M_0(4\dD\gp_{EC}\e_0) M_2(4\dD\gp_{EC}\e_0)
\nonumber \\ & \times 
\int^{2R\p/c}_{0} d\tp 
\left[ \frac{\tp c}{2R\p} - \left(\frac{\tp c}{2R\p}\right)^2 \right]
\nonumber \\ & \times 
N_e\left(\gp_{EC}; \frac{t\delta_D}{1+z}-\tp\right)\ 
\end{flalign}
where
\begin{eqnarray}
K_{EC,\delta} = \frac{\dD^6}{3\pi d_L^2} c \sT u_0 \g^{\prime 3}_{EC}\ ,
\end{eqnarray}
\begin{eqnarray}
M_1(x) = \langle y \rangle = 
\frac{\langle\e\rangle}{\g} = 
\frac{ \int^\infty_0 dy\ y\ J_{C}(x,y) }{ \int^\infty_0 dy\ J_{C}(x,y) }\ ,
\end{eqnarray}
and
\begin{eqnarray}
M_2(x) = \frac{d \ln(x)}{d\ln(xM_1(x))}\ .
\end{eqnarray}
The function $M_1(x)$ is computed using the ``Jones formula'' Compton
scattering kernel for isotropic electron and photon distributions
\citep{jones68,blumen70},
\begin{flalign}
J_C(x,y) & = 2w\ln w + (1+2w)(1-w) + 
\nonumber \\ & \times
\frac{1}{2}\frac{(xw)^2}{1+xw}(1-w)\ ,
\end{flalign}
in which
\begin{flalign}
w = \frac{y}{x(1-y)}\ .
\end{flalign} 
In Equation (\ref{flux_EC_delta}) one can find $\gp_{EC}$ from $\e$ by
solving the equation
\begin{flalign}
\label{gammaEC}
\e = \frac{\dD \gp_{EC} M_1(4\dD\gp_{EC}\e_0)}{1+z}
\end{flalign}
numerically for $\gp_{EC}$.  The function $M_1(x)$ has the asymptotes
\begin{equation}
\label{M1asympt}
M_1(x) \approx \left\{ \begin{array}{lll}
x/3 & x \ll 1 & \textrm{Thomson Regime} \\
0.691 & x \gg 1 & \textrm{Extreme KN Regime} 
\end{array}
\right. \ .
\end{equation}
This implies 
\begin{equation}
\label{gpECasympt}
\gp_{EC} \approx \left\{ \begin{array}{lll}
\frac{1}{\dD}\sqrt{\frac{3\e(1+z)}{4\e_0}} & 4\dD\gp_{EC}\e_0 \ll 1 \\
\frac{\e(1+z)}{(0.691)\dD} & 4\dD\gp_{EC}\e_0 \gg 1 
\end{array}
\right. \ 
\end{equation}
and
\begin{equation}
M_2(x) \approx \left\{ \begin{array}{lll}
1/2 & x \ll 1 & \textrm{Thomson Regime} \\
1 & x \gg 1 & \textrm{Extreme KN Regime} 
\end{array}
\right. \ .
\end{equation}

Fourier transformation of the approximate EC flux given by Equation
(\ref{flux_EC_delta}) yields
\begin{flalign}
\label{tilF_EC_delta}
\tilF^{EC,\delta}(\e,f) & = \frac{6K_{EC,\delta}(1+z)}{2\pi i t_{lc}\dD}
\tilN_e\left( \gp_{EC},\frac{(1+z)f}{\dD}\right)
\\ \nonumber & \times
\left\{ \exp\left[\frac{4\pi i f(1+z)R\p}{c\dD}\right] - 1\right\} 
\\ \nonumber & \times
M_0(4\dD\gp_{EC}\e_0) M_2(4\dD\gp_{EC}\e_0)\ .
\end{flalign}

\subsubsection{Exact Compton Cross Section}
\label{ECexact}

We follow \citet{dermer09} to calculate the flux from EC emission resulting from
the scattering of a monochromatic external radiation field that is isotropic in the
frame of the black hole and host galaxy.  We again include light travel time
effects for a spherical geometry following \citet{zacharias13}.  This
gives
\begin{flalign}
\label{EC_LC}
F^{EC}(\e,t) & = \frac{6K_{EC}(1+z)}{\dD t_{lc}} \int^{2R\p/c}_{0} d\tp 
\left[ \frac{\tp c}{2R\p} - \left(\frac{\tp c}{2R\p}\right)^2 \right]
\nonumber \\ & \times
\int_{\gp_{\min}}^{\gp_{\max}} \frac{d\gp}{\g^{\prime 2}}
N_e\p\left( \gp; \frac{t\dD}{1+z}-\tp \right)
\nonumber \\ & \times
 J_C \left( 4\dD\gp\e_0, \frac{\e(1+z)}{\gp\dD} \right)\ 
\end{flalign}
for the $\nu F_{\nu}$ EC flux.
Here 
\begin{eqnarray}
K_{EC} = \frac{3}{4}\frac{c\sT \e^2 (1+z)^2 \dD^2}{4\pi d_L^2} \frac{u_0}{\e_0^2}\ ,
\end{eqnarray}
where
\begin{eqnarray}
\gp_{\min} = \frac{\e(1+z)}{2\dD}\left( 1 + \sqrt{1 + \frac{1}{\e\e_0(1+z)}}\right)\ ,
\end{eqnarray}
and
\begin{eqnarray}
\gp_{\max} = \g_2\ .  
\end{eqnarray}
Recall that $\g_2$ denotes the high-energy cutoff in the electron
injection spectrum (see Equation [\ref{tilQ}]).  The
Fourier-transformed EC flux is then given by
\begin{flalign}
\label{tilF_EC}
\tilF^{EC}(\e,f) & = \frac{6K_{EC}(1+z)}{2\pi if t_{lc}\dD}
\left\{ \exp\left[ \frac{4 \pi i f(1+z) R^\prime}{c\dD}\right] - 1 \right\}\
\nonumber \\ & \times 
\int_{\gp_{\min}}^{\gp_{\max}} \frac{d\gp}{\g^{\prime 2}}
\tilN_e\p\left( \gp; \frac{(1+z)f}{\dD} \right) 
\nonumber \\ & \times 
J_C\left( 4\dD\gp\e_0, \frac{\e(1+z)}{\gp\dD} \right)\ .
\end{flalign}

\subsection{Numerical Results}

As described in Paper I, for compact sources, the light travel time
effect (Appendix \ref{delayappendix}) will not be noticeable.  In this
section, we remove this effect to more easily display the comparison
between the $\delta$ approximation and the full calculation.  Without
light travel time effects (valid for $R^\prime_b \ll
c\dD[f(1+z)]^{-1}$), the synchrotron PSDs in the $\delta$-function
approximation (Section \ref{syndeltaapprox}) and full calculation
(Section \ref{synexact}), respectively, can be written as
\begin{flalign}
\label{PSDsydelta}
S^{sy,\delta}(\e,f) & = \left(\frac{6K_{sy,\delta}(1+z)}{\dD}\right)^2
\left| \tilN_e\left(\gp_{sy}, \frac{(1+z)f}{\dD}\right) \right|^2 \ ,
\end{flalign}
and
\begin{flalign}
\label{PSDsy}
S^{sy}(\e,f) & = \left(\frac{6K_{sy}(1+z)}{\dD}\right)^2
\nonumber \\ & \times 
\Biggr| \int_1^{\infty} d\gp \tilN_e\p \left(\gp; \frac{(1+z)f}{\dD}\right) 
\nonumber \\ & \times 
R_{CS}\left(\frac{2\e(1+z)}{3\e_B\dD\g^{\p 2}} \right)\Biggr|^2 \ .
\end{flalign}

\begin{figure}
\vspace{2.2mm} 
\epsscale{0.9} 
\plotone{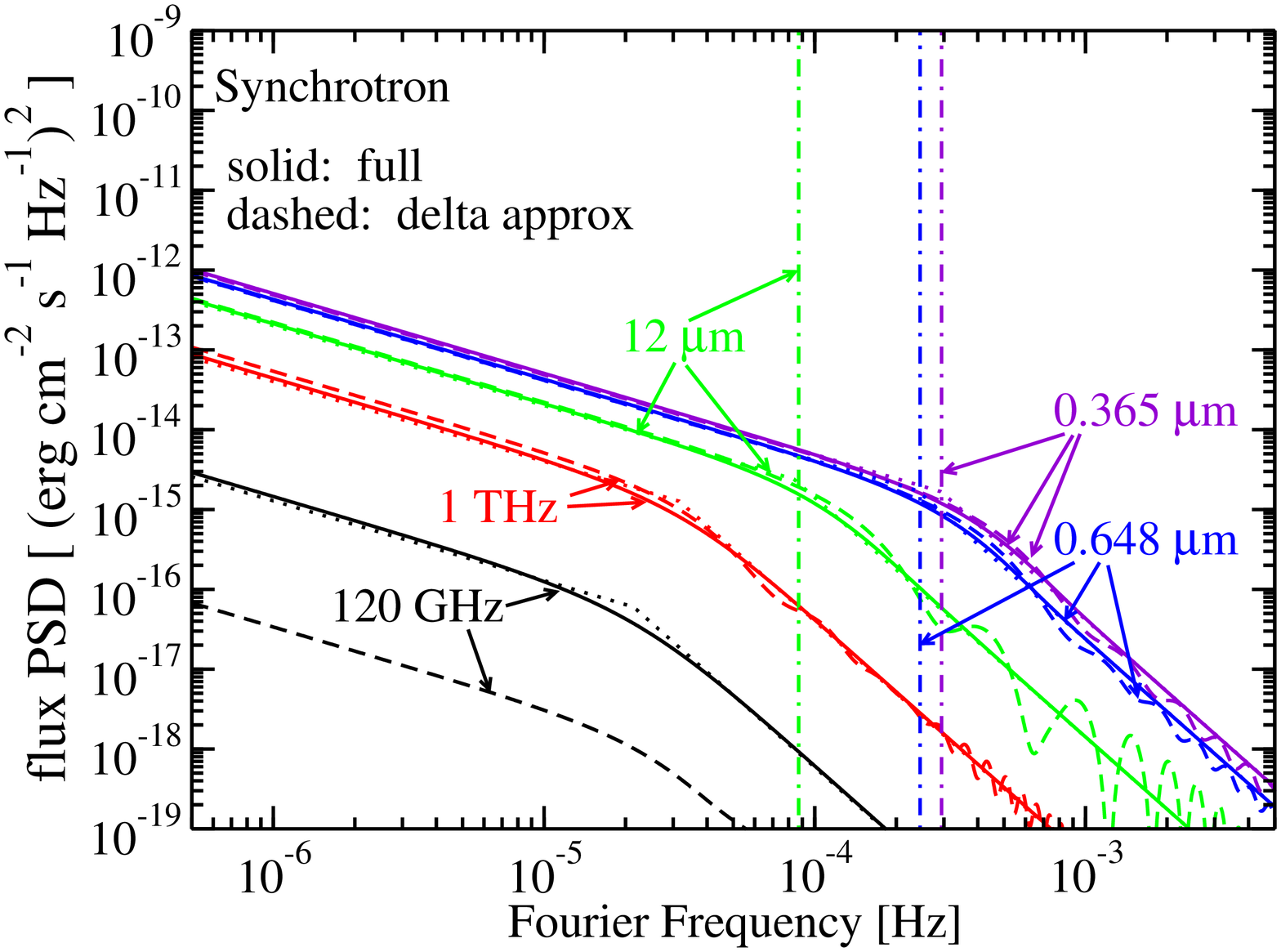}
\caption{The synchrotron PSD for the $\delta$-function approximation
(dashed curves; Equation [\ref{PSDsydelta}]), full calculation (solid
curves; Equation [\ref{PSDsy}]), and broken power-law fit (dotted
curves; Equation).  Parameters are $q=2$, $a=1$,
$\tp_{\rm esc}=10^5\ \s$, $\langle L_{\rm inj}\rangle=10^{42}\ \erg\
\s^{-1}$, $\Delta t=1$\ year, $\g_1=10^2$, $\g_2=10^5$, $R\p=10^{15}\
\cm$, $B=1$\ G, $\G=\dD=30$, $u_0=10^{-3}\ \erg\ \cm^{-3}$,
$\e_0=5\times10^{-7}$, and $z=1$.  At this redshift with cosmology
($h$, $\Omega_m$, $\Omega_\Lambda$)=(0.7, 0.3, 0.7) the luminosity
distance $d_L=2\times10^{28}\ \cm$.  The observed photon frequency
($m_ec^2\e/h$) or wavelength ($hc/[m_ec^2\e]$) is shown.
Dashed-dotted lines indicate $f=(3t_{\rm cool})^{-1}$ for each curve
where $t_{\rm cool}<t_{\rm esc}$.  }
\label{syPSDfig}
\vspace{2.2mm}
\end{figure}

A comparison of synchrotron PSDs computed the two different ways is
shown in Figure \ref{syPSDfig}.  Agreement between the
$\delta$-function approximation and the full calculation is good,
except for the 120 GHz curve.  For this frequency, photons come from
electrons with $\gp < \g_1$.  Clearly for this range the
$\delta$-function approximation is not accurate.  With an
observer-frame cooling timescale defined by
\begin{flalign}
\label{cooltime}
t_{\rm cool}(\e) = \frac{1+z}{\dD} \int_\g^{\g_2} \frac{d\gp}{|\dot{\g}(\gp)|}\ ,
\end{flalign}
the PSDs with $t_{\rm cool}<t_{\rm esc}$ exhibit breaks at
$f=[3t_{\rm cool}(\e)]^{-1}$.  As $t_{\rm cool}$ gets closer and closer to
$t_{\rm esc}$, and eventually gets longer than $t_{\rm esc}$, this
approximation breaks down.  For synchrotron, $\g$ in Equation
(\ref{cooltime}) is calculated from $\e$ using Equation
(\ref{gamma_sy}).  The full curves ``wash out'' the many minima of the
$\delta$-function approximation, so that the PSDs more closely
resemble broken power-laws.  The break has a magnitude of 2, i.e., the
break is from $S(\e,f)\propto f^{-a}$ to $S(\e,f)\propto f^{-(a+2)}$.

\begin{figure}
\vspace{2.2mm} 
\epsscale{0.9} 
\plotone{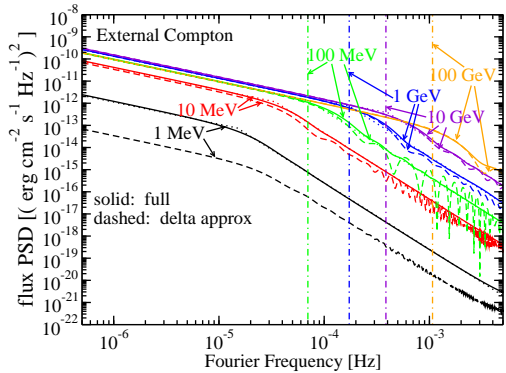}
\caption{The EC PSD for the $\delta$-function approximation (dashed
curves; Equation [\ref{PSDECdelta}], full calculation (solid curves;
Equation [\ref{PSDEC}]), and broken power-law fit (dotted curves).
Parameters are the same as in Figure \ref{syPSDfig}.  The observed
photon energy ($m_ec^2\e$) is shown.  Dashed-dotted lines indicate
$f=(3t_{\rm cool})^{-1}$ for each curve where $t_{\rm cool}<t_{\rm
esc}$.  }
\label{ECPSDfig}
\vspace{2.2mm}
\end{figure}

Without light travel time effects ($R^\prime_b \ll
c\dD[f(1+z)]^{-1}$), the EC PSDs in the $\delta$-function
approximation (Section \ref{ECdeltaapprox}) and full calculation
(Section \ref{ECexact}), respectively, can be written as
\begin{flalign}
\label{PSDECdelta}
S^{EC,\delta}(\e,f) & = \Biggr[\frac{6K_{EC,\delta}(1+z)}{\dD}
\nonumber \\ & \times 
M_0(4\dD\gp_{EC}\e_0) M_2(4\dD\gp_{EC}\e_0) \Biggr]^2
\nonumber \\ & \times 
\left| \tilN_e\left(\gp_{EC}, \frac{(1+z)f}{\dD}\right) \right|^2 \ ,
\end{flalign}
and
\begin{flalign}
\label{PSDEC}
S^{EC}(\e,f) & = \left(\frac{6K_{EC}(1+z)}{\dD}\right)^2
\nonumber \\ & \times 
\Biggr| \int_{\gp_{\min}}^{\gp_{\max}} \frac{d\gp}{\g^{\prime 2}}
\tilN_e\p\left( \gp; \frac{(1+z)f}{\dD} \right) 
\nonumber \\ & \times 
J_C\left( 4\dD\gp\e_0, \frac{\e(1+z)}{\gp\dD} \right)\Biggr|^2 \  .
\end{flalign}
Figure \ref{ECPSDfig} plots the EC PSDs computed the two different
ways.  Here for EC, $\gamma$ in Equation (\ref{cooltime}) is found
from $\e$ by solving Equation (\ref{gammaEC}) numerically.  Minima
features in the $\delta$-function approximation are again washed out
in the full calculation, and agreement is good, except for the 1 MeV
curve, where photons are generated primarily from electrons with $\gp
< \g_1$.  As with the synchrotron PSDs, in the EC PSDs
for $t_{\rm cool}<t_{\rm esc}$, breaks are seen at $f=[3t_{\rm
cool}]^{-1}$, and again this approximation is less valid as $t_{\rm
cool}$ approaches $t_{\rm esc}$.

\section{Time Lags}
\label{timelag}

\begin{figure}
\vspace{2.2mm} 
\epsscale{0.9} 
\plotone{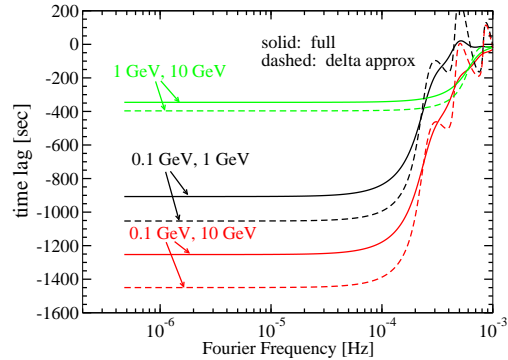}
\caption{EC flux time lags.  Solid curves show the lags computed with
the full calculation (Equation [\ref{tilF_EC}]), dashed curves show
the lags computed with the $\delta$-function approximation (Equation
[\ref{tilF_EC_delta}]).  }
\label{fluxlag}
\vspace{2.2mm}
\end{figure}

Time lags as a function of Fourier frequency are given by
\begin{flalign}
\label{deltaT}
\Delta T(\e_a, \e_b, f) = \frac{1}{2\pi f}
\arctan\left[ \frac{Y_I(\e_a, \e_b, f)}{Y_R(\e_a, \e_b, f)}\right]
\end{flalign}
(e.g., Paper I) where $Y_R(\e_a, \e_b, f)$ and $Y_I(\e_a, \e_b,
f)$ are defined by
\begin{flalign}
\label{complexconj}
\tilF(\e_a,f)\tilF^*(\e_b,f) = Y_R(\e_a, \e_b, f) + i\ Y_I(\e_a, \e_b, f)\ .
\end{flalign}
Equations (\ref{deltaT}) and (\ref{complexconj}) above are combined
with Equations (\ref{tilF_EC_delta}) and (\ref{tilF_EC}) to compute
the time lags with the Moderski approximation and full
calculation for the EC time lags.  The results are shown in Figure
\ref{fluxlag}.  Note that light travel time effects will play no part
in time lags, since they are energy-independent.  
Our numerical results indicates that the time lags in the
limits $f\ll [2\pi t_{\rm cool}(\e_b)]^{-1}$ and $f\ll [2\pi t_{\rm
cool}(\e_b)]^{-1}$, are
\begin{flalign}
\Delta T(\e_a,\e_b,f) \approx \left[ t_{\rm cool}(\e_b) - t_{\rm cool}(\e_a) \right]/3\ 
\end{flalign}
for the full calculation, where $t_{\rm cool}(\e)$ is the new definition
from this paper, Equation (\ref{cooltime}).  The Moderski $\delta$
function approximation however is about 15\% higher in all cases.
This is also true for synchrotron time lags, which are not shown.
Clearly the $\delta$ function approximations are not as accurate for
time lags as they are for PSDs.  

In this paper, we use the standard convention that $\e_b < \e_a$
\citep[e.g.,][]{kroon14}, so that positive lags refer to hard lags
(i.e., the hard channel lags behind the soft channel), and negative
lags to soft lags (i.e., the soft channel lags behind the hard
channel).  Thus, Figure \ref{fluxlag} shows our model only reproduces
soft lags.  Note that this is in contrast to Paper~I, where we did not
use this convention.  In X-ray observations of BL Lac objects, both
soft lags \citep[e.g.][]{zhang02_pks2155} and hard lags
\citep[e.g.,][]{zhang02_mrk421} have been observed.  Our model can
reproduce the soft lags but not the hard lags, contrary to our
discussion in Section 6.2 of Paper~I.  This is because the continuity
equation treated here only includes losses (due to synchrotron and EC
emission) and does not include any energization processes.  Inclusion
of acceleration effects in our model may be able to
explain the hard lags if the acceleration timescale is shorter than
the cooling timescale, as suggested by \citet{zhang02_mrk421}.  We
will explore this possibility in future work (Lewis, Becker, \& Finke
2015, in preparation).

\section{Comparison with Observations}

In Section 6 of Paper I, we compared our model with several PSD
observations.  In this section, we update some of our conclusions,
based on the improvements to our model.  Results from Paper I not
described below are unchanged.

\subsection{Oscillatory Features}

The use of the accurate expressions eliminate the oscillatory
structures in the PSDs seen with the $\delta$-approximations, so that
the PSDs more closely resemble broken power-laws.  Possible minima
from observed PSD, such as the VHE PSD of PKS 2155$-$304 presented by
\citet{aharonian07_2155}, or the QPOs observed by
e.g. \citet{Lachowicz09}, \citet{gupta09} or \citet{rani10}, we now
believe are unlikely to be associated with the minima features predicted
in Paper~I.  These features seem to be an artifact of the
$\delta$-function approximation and they are not present in more
accurate calculations.

\subsection{The $\g$-ray PSD of 3C~454.3}
\label{gray3c454}

As reported by \citet{nakagawa13}, the LAT $\g$-ray PSD of 3C 454.3
has a break at a frequency of $1.5\times10^{-5}$\ Hz \citep[although
see][]{sobol14}.  If at LAT energies $t_{\rm cool}<t_{\rm esc}$, this
implies a cooling timescale of $2.3\times10^5$\ s in the observer
frame.  For parameters $\G=\dD=30$, $B=1.0\ \Gauss$, $u_0=10^{-4}\
\erg\ \cm^{-3}$, $\g_2=10^5$, and assuming the LAT is dominated by
photons at 100 MeV, one can numerically solve for the seed photon
energy, and obtain $\e_0=1.8\times10^{-5}$, which is quite close to
the value for Ly$\alpha$ ($2.0\times10^{-5}$ in $m_ec^2$ units).  For
the same set of parameters, except $u_0=10^{-2}\ \erg\ \cm^{-3}$, one
gets $\e_0=7.0\times10^{-4}$ (or $0.36\ \keV$), and for $u_0=10^{-1} \
\erg\ \cm^{-3}$, one gets $\e_0=3.3\times10^{-3}$ (or $1.7\ \keV$).
The latter two are unrealistically high for seed photon energies, and
Ly$\alpha$ does make more sense as a seed photon source, since it is
the most prominent line seen in optical spectra
\citep[e.g.,][]{telfer02}, however this is an unrealistically low
energy density.  In this regard, the Klein-Nishina effects do not
alter our conclusions from Paper I.

\subsection{The Optical PSDs of blazars}

\citet{edelson13} have used {\em Kepler} observations to identify a
``bend'' in the PSD of the BL Lac W2R1926+42 corresponding to a
period of 4 hours.  Assuming the emission detected by {\em Kepler}
is dominated by detections at 5000 \AA, and for similar parameters as
in Section \ref{gray3c454}, with $u_0=10^{-4}\ \erg\ \cm^{-3}$ one
gets $\e_0=2.3\times10^{-7}$, about what one would expect for
scattering of dust torus emission; for $u_0=10^{-2}\ \erg\ \cm^{-3}$
one gets $\e_0=2.5\times10^{-5}$, about what one expects for
scattering Ly$\alpha$ photons; and for $u_0=10^{-1}\ \erg\ \cm^{-3}$
one gets $\e_0=1.2\times10^{-4}$.  The latter is probably an
unrealistic value for $\e_0$, and the values of $u_0$ for the former
are about what one would expect for these seed photon sources,
although for Ly$\alpha$ the energy density is a bit low.  Thus for
this source, scattering of dust torus emission seems the most likely,
although this conclusion is strongly dependent on the assumed
parameters.  In Section \ref{seedobtain} we describe a method of
determining the seed photon source that is more model-independent.

\section{Method for Determining Seed Photon Energy}
\label{seedobtain}

Here we outline a technique for determining the energy of the seed
photon source for EC.  We assume the external seed photon source can
be approximated as isotropic in the galaxy's frame, and that it is
monochromatic with dimensionless energy $\e_0$.  We also assume from
the breaks or time lags in the PSDs of blazars at several energy
ranges in EC one can obtain the cooling timescale, as described in
Sections \ref{observedpsd} and \ref{timelag}.  For example, one might
find breaks in {\em Fermi}-LAT PSDs at $m_ec^2\e_a=0.1$\ GeV,
$m_ec^2\e_b=1.0$\ GeV, and $m_ec^2\e_c=10.0$\ GeV, and thus
the cooling timescales at these energies.

If one is observing synchrotron or EC, one can use Equation
(\ref{cooltime}) to compute the observer frame cooling timescale from
synchro-Compton losses, which we rewrite as
\begin{flalign}
t_{\rm cool}(\e) = \frac{3(1+z)m_e c^2 \e_0}{c\sT u_B}
\int^{x_2}_{x_1} \frac{dx}{x^2}\frac{1}{1 + A_C M_0(x)} \ 
\end{flalign}
where
\begin{flalign}
x_2 = 4\G\g_2\e_0
\end{flalign}
and
\begin{eqnarray}
A_C = \frac{\G^2 u_{0}}{u_B}\ 
\end{eqnarray}
is the Compton dominance, assuming $\dD=\G$.
If the cooling timescale is estimated from a PSD or time lag that is emitting
synchrotron, the integral's lower limit is
\begin{flalign}
x_{1,sy} = 4\e_0\left(\frac{\e(1+z)\dD}{\e_B}\right)^{1/2}\ .
\end{flalign}
If the cooling timescale is estimated from EC, then the integral's lower limit is
\begin{flalign}
x_{1,EC} & = 4\G\gp_{EC}\e_0 = \frac{4\e\e_0(1+z)}{M_1(x_{1,EC})} 
\nonumber\\ 
& \approx \left\{ \begin{array}{lll}
2\sqrt{3\e\e_0(1+z)} & 4\G\gp_{EC}\e_0 \ll 1 \\
4\e\e_0(1+z)/(0.691) & 4\G\gp_{EC}\e_0 \gg 1 
\end{array}
\right. \ .
\end{flalign}
For computing $x_1$ above we have made use of the $\delta$ function
approximations for synchrotron and EC, as described in Section
\ref{syndeltaapprox} and \ref{ECdeltaapprox}.  For EC, a function 
created from three cooling timescales, 
\begin{flalign}
\label{coolfunction}
r(\e_a,\e_b,\e_c) = \frac{t_{\rm cool}(\e_a) - t_{\rm cool}(\e_c)}{t_{\rm cool}(\e_a) - t_{\rm cool}(\e_b)}
\end{flalign} 
is dependent only on $\e_0$ and $A_C$.  In principle, $A_C$ can be
determined from the broadband spectral energy distribution
\citep[e.g.,][]{meyer12,finke13}.  For synchrotron, this function is
additionally dependent on the ratio $\delta_D/\e_B$.  Observations of
FSRQ synchrotron PSDs and lags will also suffer from contamination by
thermal emission, from the accretion disk and dust torus.  Therefore
we will only be concerned with the $\gamma$-ray emission from EC.

\begin{figure}
\vspace{2.2mm} 
\epsscale{0.9} 
\plotone{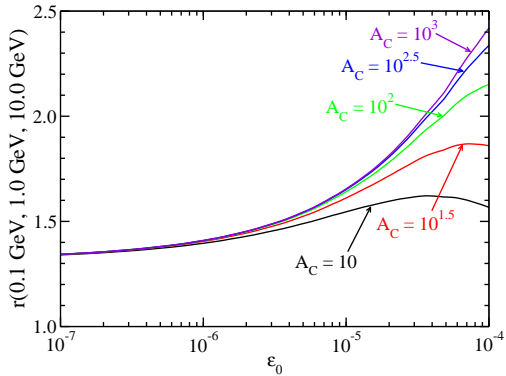}
\caption{The function $r(0.1\ \GeV, 1.0\ \GeV, 10.0\ \GeV)$ from
Equation (\ref{coolfunction}) plotted as a function of dimensionless
seed photon energy $\e_0$ for different values of Compton dominance
$A_C$.}
\label{coolratio_01}
\vspace{2.2mm}
\end{figure}

We plot the $r(0.1\ \GeV, 1.0\ \GeV, 10.0\ \GeV)$ for different values
of $A_C$ in Figure \ref{coolratio_01}.  These plots demonstrate that
given $A_C$ and $t_{\rm cool}$ for three energies, one can estimate the
energy of the seed photon source, $\e_0$.  For low values of $\e_0$
($\e_0\la10^{-6}$), the scattering at all three energies will be in
the Thomson regime, and the value will be mostly independent of $\e_0$
or $A_C$.  Only at higher values of $\e_0$ will Klein-Nishina effects
become apparent.

Although in principle it may be possible to determine $\e_0$ from PSDs
or time lags computed from {\em Fermi}-LAT light curves, in practice
this will be extremely difficult.  During the brightest
flares, it is possible to probe timescales $\sim$\ 1 hour
\citep[e.g.,][]{tavecchio10,abdo11_3c454.3,brown13,saito13,nalewajko13},
but this is rare.  More typically, integrations over a few days to a
week are needed to significantly detect a source with the LAT
\citep[e.g.,][]{abdo10_var}, so that probing the $\sim$ 1 hour
timescales necessary to detect the breaks is unlikely in most cases.
So far, three FSRQs have published spectra from imaging atmospheric
Cherenkov telescopes (IACTs)
\citep{albert08_3c279,aleksic11,abramowski13_1510}, and one more has
been detected by IACTs \citep{mirzoyan15,mukherjee15}. It is certainly
reasonable to expect that the proposed Cherenkov Telescope Array
\citep[CTA;][]{actis11} will be able to get significant detections
with shorter integration times than the LAT, and therefore it may be
able to probe these timescales.  However, at higher energies, the
cooling timescales will also be shorter, making the breaks more
difficult to observe.  If the breaks are on sub-hour timescales, a
time baseline (i.e., $\Delta t$ in our notation) of about 10 times the
timescale of the expected break should be sufficient, so that a single
night of observing would probably be a long enough baseline to observe
a break.  Probing short enough timescales would be a greater issue.
Further, any source detected at $\ga20\ \GeV$ is almost certainly
making $\g$ rays outside the BLR, otherwise the $\g$-rays would be
attenuated by $\g\g$ absorption with the Ly$\alpha$ photons
\citep[e.g.,][]{stern14}.  However, an analysis with CTA could verify
our results.  

The technique outlined here is similar to obtaining the cooling
timescales by modeling the decay times of individual flares, as
discussed by \citet{dotson12}, although our model has a few
advantages.  For example, with the new method, one is able to take
advantage of a long series of data, rather than individual flares,
possibly leading to more statistically significant results.  When
modeling individual flares, it is not clear if the decrease in flux
from a flare is due to radiative cooling or due to the rapid decrease
in the electron acceleration rate.  The disadvantage of our technique
is it may be that different flares from the same object have different
origins.  Perhaps some flares occur close to the BH, so that the seed
photon source is from the BLR, while others occur farther away, so
that the seed photon source is the dust torus.  PSDs based on long
timescale light curves would have a difficult time disentangling these
cases.

\section{Discussion}

We have expanded upon our previous theoretical scenario for computing
PSDs and time lags (Paper~I).  We have included the full Compton cross
section in our solution to the Fourier-transformed continuity
equation.  As in Paper~I, we find breaks in the resulting PSDs
associated with the cooling timescale, however, we must revise our
definition of the cooling timescale.  We have compared the simple
$\delta$-function approximations for synchrotron and Compton emission
to more accurate expressions.  For PSDs the agreement between the
approximations and the accurate expressions is very good, except for
photons produced primarily by electrons with $\gp < \gp_1$.  As in
Paper~I, we find that the breaks in synchrotron and EC PSDs always
have magnitudes of 2, i.e., the break is from $S(\e,f)\propto f^{-a}$
to $S(\e,f)\propto f^{-(a+2)}$.

At low Fourier frequencies, time lags calculated with the
$\delta$-approximations give lags about 15\% greater than lags
computed with the accurate expressions.  The cause of the deviation is
not clear.  However, as in Paper~I, we confirm that the time lags at
low Fourier frequencies can be associated with the difference in
cooling timescales of the two energy channels.

Based on our theoretical work, it seems that one could in principle
measure the cooling timescale based on observations of blazar PSDs or
time lags.  If one can measure the cooling timescale from
$\gamma$-rays produced by EC at three different energies, while also
determining the Compton dominance from the blazar's broadband SED,
then in principle determine the energy of the seed photon source.
This would give a strong indication as to the location of the
$\gamma$-ray emitting region, something which has so far eluded
understanding.  Such a feat is unlikely to be possible with currently
operating $\gamma$-ray telescopes such as the {\em Fermi}-LAT, but may
be possible with CTA.

\acknowledgements 

We are grateful to the referee for helpful comments that have
improved this manuscript, and to C.\ Dermer and S.\ Larsson for
useful discussions.  JDF is supported by the Chief of Naval Research.

\appendix

\section{Light Travel Time Delay}
\label{delayappendix}

\begin{figure}
\vspace{2.2mm} 
\epsscale{0.6} 
\plotone{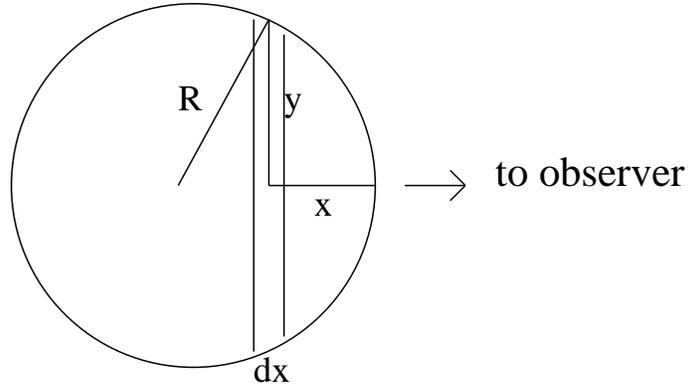}
\caption{Geometric sketch of spherical emitting blob for the
purpose of computing light travel time effects.  This sketch is in the 
frame co-moving with the blob.  }
\label{varfig}
\vspace{2.2mm}
\end{figure}

In this section we derive the light travel time effect, directly
following \citet[][c.f.\ Appendix C of Paper~I]{zacharias13}.  The
spherical emitting region with radius $R$, volume $V=4\pi R^3/3$ is
homogeneous, with variations taking place throughout the sphere
simultaneously.  Its total emitting flux is $F$ at time $t$, but the
observer will see light from the closer part of the sphere arrive
before the light from the farther part of the sphere.  For the
calculation of this effect, the emitting region is divided into an
infinite number of ``slices'' each with infinitesimally small
thickness $dx$ emitting flux $dF$ at time $t$.  The slices are cut
perpendicular to the direction of the observer so that each one has
radius
\begin{flalign}
y=\sqrt{ R^2 - (R-x)^2 }\ ,
\end{flalign}
cross sectional area
\begin{flalign}
A = \pi y^2 = \pi(2Rx-x^2)\ ,
\end{flalign}
and volume
\begin{flalign}
dV = A\ dx 
\end{flalign}
(see Figure \ref{varfig}).  The flux emitted by each slice as a
fraction of the whole is proportional to the volume of the slice as a
fraction of the whole, i.e.,
\begin{flalign}
\frac{dF}{F} = \frac{dV}{V}\ .
\end{flalign}
The observer sees flux from each slice delayed by time $t=x/c$ 
so that the observed flux at time $t_{obs}$ is
\begin{flalign}
\label{flux1}
F_{obs}(t_{obs}) & = \int dF(t_{obs}-t) = \int F(t_{obs}-t)\ \frac{dV(t_{obs})}{V}\ .
\nonumber \\
                 & = \frac{3c}{R} \int^{2R/c}_0 dt\ F(t_{obs}-t) 
                     \left[ \frac{tc}{2R} - \left(\frac{tc}{2R}\right)^2 \right]\ .
\end{flalign}
Combining this with standard formulae for synchrotron and Compton
emission for $F(t)$ \citep[e.g.,][]{dermer09_book} leads to the
results in Sections \ref{synchsection} and \ref{comptonsection}.


\bibliographystyle{apj}
\bibliography{variability_ref,EBL_ref,references,mypapers_ref,blazar_ref,sequence_ref,SSC_ref,LAT_ref,3c454.3_ref}

\end{document}